\documentclass[a4paper,english]{IEEEtran}
\usepackage[T1]{fontenc}
\usepackage[latin9]{inputenc}
\usepackage{color}
\usepackage{babel}
\usepackage{verbatim}
\usepackage{url}
\usepackage{graphicx}
\usepackage[pdfusetitle,
 bookmarks=true,bookmarksnumbered=false,bookmarksopen=false,
 breaklinks=false,pdfborder={0 0 1},backref=false,colorlinks=true]
 {hyperref}

\makeatletter

\pdfpageheight\paperheight
\pdfpagewidth\paperwidth

\makeatother

\begin{document}
\title{RF Fingerprinting of DSP Implementations using Machine Learning: An
Over-the-Air Case Study}
\author{Stefan Scholl\\
research@panoradio-sdr.de}
\maketitle
\begin{abstract}
Radio frequency fingerprinting aims to identify transmitters by exploiting
small differences in the emitted signal waveform. Common approaches
exploit hardware-dependent imperfections for that purpose. This work
instead investigates whether differences between digital signal processing
implementations of a transmitter can be identified from their radiated
signals. As a practical case study, five different software-based
DSP implementations for generating a 2-FSK RTTY communication waveform
are considered. In a monitoring receiver, a classifier can identify
these DSP implementations using convolutional neural networks that
are trained on augmented software-generated signals represented as
raw IQ, power spectrum or instantaneous frequency. The trained networks
are evaluated on over-the-air recordings obtained using a shortwave
transceiver and a software-defined radio receiver. The best neural
network combines IQ and the power spectrum representation and achieves
an accuracy of up to 90\%. The study further examines the influence
of the signal SNR and the observation bandwidth on the classification
accuracy. The results demonstrate that DSP implementations leave distinguishable
characteristics in the transmitted signals and suggest that DSP-dependent
fingerprinting can provide complementary information for specific
emitter identification systems.
\end{abstract}


\section{Introduction}

\subsection{Radio Frequency Fingerprinting for Specific Emitter Identification}

Specific Emitter Identification (SEI) is a task in radio signal monitoring
that focuses on the recognition of transmitters rather than the transmitted
modulation or signal type or mode. Consequently, SEI can distinguish
between different transmitters that emit the same signal type or mode.
In contrast to signal identification and automatic modulation recognition,
which focus on determining the characteristics of the transmitted
signal, SEI seeks to exploit device-specific characteristics that
are linked to the specific transmitters. %

\begin{figure}
\centering
\includegraphics[width=0.8\columnwidth]{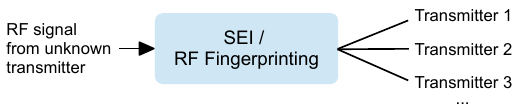}

\caption{\label{fig:Specific-Emitter-Identification}Specific emitter identification
recognizes the transmitters.}

\end{figure}

SEI is a classification problem with the input being an RF signal
from an unknown transmitter and the output classes corresponding to
the different transmitter devices, as shown in Figure \ref{fig:Specific-Emitter-Identification}.
Since the signal characteristics can be very similar, SEI is more
challenging than modulation recognition and signal identification.
Therefore, in some cases, it is only possible to identify groups of
transmitters, such as those from a common manufacturer or from a specific
production series, instead of being able to properly separate individual
transmitters.

SEI can be implemented through RF fingerprinting. It exploits subtle
variations in transmitter implementation that result in slightly different
output signal waveforms, which can be used to identify individual
transmitters. These variations can, in general, originate from different
parts of the transmitter, such as:
\begin{itemize}
\item Analog electronics, e.g. circuit design, PCB design and shielding,
parts selection, manufacturing tolerances
\item Digital-to-analog converter (DAC) properties
\item Digital signal processing (DSP) algorithms, e.g. filter design, quantization,
sample rates
\end{itemize}
Typical resulting signal characteristics that can be used for RF fingerprinting
are, e.g.:
\begin{itemize}
\item Details of the spectral shape (e.g. originating from filters)
\item Spurious emissions (e.g. from oscillators and non-linear blocks)
\item Signal transients
\item Frequency and clock drifts
\item Phase noise properties
\end{itemize}
Recognition algorithms for RF fingerprinting exploit these subtle
differences in the emitted signals to identify transmitters.

\subsection{Common Approaches to RF Fingerprinting}

Classical RF fingerprinting typically follows a two-step approach:
First, characteristic features are extracted from the received signal
to emphasize signal differences between transmitters. These features
are often manually designed by an expert and can include statistical
properties of the signal, higher-order moments and cumulants, spectral
characteristics, or transient signal characteristics. Second, a classifier
is applied to these features to determine the transmitter class. Commonly
used classifiers include support vector machines, k-nearest neighbors,
decision trees, or other statistical classification methods \cite{Yuan.2014}.

More recent approaches rely on deep learning algorithms, such as convolutional
neural networks (CNNs), LSTMs and transformers, which can operate
on raw signal representations, such as IQ, or spectral data \cite{Wu.2018,Jian.2020}.
In contrast to classical approaches, these methods do not require
the manual design of features, but automatically learn the characteristic
signal properties from a large amount of training data during a training
process.

For a comprehensive overview of classical and recent techniques for
RF fingerprinting, the interested reader is referred to \cite{AnuJagannath.2022,Xie.2023}
and \cite{Yan.2025}.

\subsection{Scope and Contributions}

This work focuses on RF fingerprinting based on properties of DSP
implementations rather than on imperfections of analog hardware components.
It investigates how the characteristics of radio DSP implementations
can be used for RF fingerprinting. More specifically, this work
\begin{itemize}
\item investigates whether algorithms for waveform generation leave classifiable
DSP-dependent characteristics in the transmitted signal,
\item applies a deep learning approach that trains neural networks exclusively
on augmented software-generated data and evaluates on over-the-air
recordings,
\item compares raw IQ, spectral, instantaneous-frequency, and combined signal
representations,
\item analyzes the influence of SNR and receiver bandwidth on the classification
accuracy.
\end{itemize}
The remainder of the paper is organized as follows: Section \ref{sec:RF-Fingerprinting-based-on-dsp}
presents the DSP-based RF fingerprinting approach and discusses DSP
effects that can influence the exact shape of the output waveform.
Section \ref{sec:Case-Study:-Amateur} presents a case study on amateur
radio signals, including the transmission system, training and test
datasets, machine-learning approach, and classification results. Section
\ref{sec:Summary} concludes the paper.

\section{\label{sec:RF-Fingerprinting-based-on-dsp}RF Fingerprinting based
on DSP Implementation}

\begin{figure}
\begin{centering}
\includegraphics[width=1\columnwidth]{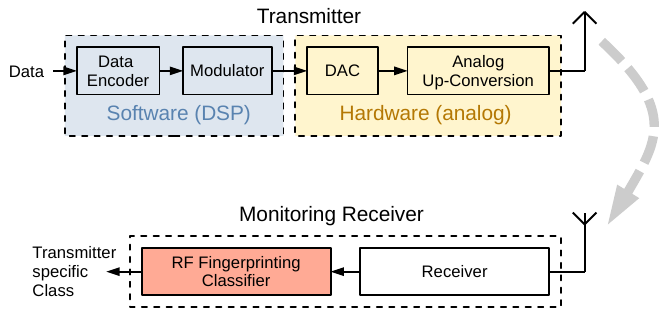}
\par\end{centering}
\caption{\label{fig:hardware-setup}Exemplary setting of a typical transmitter
and a monitoring receiver using RF fingerprinting}

\end{figure}

A typical transmitter consists of a combination of DSP, DAC and analog
hardware as depicted in Figure \ref{fig:hardware-setup}. DSP algorithms
are often used to generate the transmission waveform and perform encoding,
framing and baseband modulation of the data to be transmitted. DSP
algorithms can be implemented as software running on a processor or
as a digital circuit implemented on an FPGA or ASIC. The digital signal
is then converted to an analog signal using a DAC. Finally, the analog
signal is up-converted, amplified and filtered by analog hardware
before being transmitted via an antenna. Note that the division between
digital and analog signal processing may vary and additional functions
may be implemented in DSP, such as up-conversion.

Transmitter-specific signal properties can be introduced at different
parts of the transmitters, namely in implementation details of DSP
algorithms, DAC and RF hardware. While most of the literature focuses
on analog hardware imperfections, this paper focuses on the influence
of the DSP algorithms for generating the modulated signal. Therefore,
the approach in this work aims to differentiate transmitters based
on their DSP implementation, rather than the hardware implementations.
Note that DSP-based features alone do not necessarily identify an
individual physical device, especially when two transmitters use the
exact same DSP implementation and underlying digital hardware, but
they can provide a valuable component of a larger SEI system.

The waveform-generating DSP algorithms create the modulated signal
in the digital domain and are implemented in software or as a digital
circuit. This includes the encoding of the data to be transmitted,
mapping of the bits to modulation symbols and generating the time-domain
baseband waveform. This process may also include spectral filtering,
sample rate conversion and a first up-conversion.

The exact shape of the output signal depends on the details of the
DSP implementation. The following design aspects can influence the
output waveform:
\begin{itemize}
\item Modulation shaping
\item Digital filtering
\item Digital mixing and numerically controlled oscillators
\item Digital automatic gain control
\item Numerical quantization and rounding
\item Sample rates and digital clocks
\item Sample rate conversion
\item Implementation-specific deficiencies, such as insufficient suppression
of spurious components
\end{itemize}
These implementation details may depend on the manufacturer, developer,
design style, algorithm selection, performance requirements and available
hardware (e.g. CPU, GPU, FPGA, microcontroller).

\section{\label{sec:Case-Study:-Amateur}Case Study: Amateur Radio Communication}

\subsection{Overview of the Approach}

The DSP-based RF fingerprinting approach is now investigated in a
practical case study using amateur radio communication signals. It
considers transmitters with different waveform-generation software
that output the same transmission mode, but use slightly different
DSP implementations to generate the signals. %
{} Since the differences can be subtle and the influence of these effects
is difficult to predict, a data-driven approach is applied using machine
learning with deep neural networks. For that purpose, first, training
data is generated from different DSP implementations. With that data,
several CNNs are trained to classify signals using different signal
representations. Finally, the trained models are evaluated using over-the-air
recordings obtained from a real-world transmitter and receiver setup.
This approach enables the assessment of the trained machine learning
systems under real-world conditions, rather than relying only on synthetic
data, which would limit the validity of the evaluation.

\subsection{Digital communication with 2-FSK RTTY}

As a practical case study, a 2-FSK digital communication system is
considered that is known as radio teletype (RTTY). This type of communication
is widely used in narrow-band shortwave communications, but can also
be found in the VHF or UHF bands. RTTY encodes characters using the
Baudot code and uses frequency-shift keying modulation with two tones
(2-FSK), where binary symbols are represented by two discrete RF frequencies.
RTTY is employed, e.g., in commercial and military radiotelegraphy,
maritime communications, aviation weather broadcasts, and amateur
radio. 

This work focuses on amateur radio signals, because for this application
DSP implementations for generating the waveforms are readily available
and amateur radio allows experimental over-the-air communication links
to be established easily. In contrast, the implementations of professional
systems are typically proprietary, which makes their scientific investigation
more difficult. Nevertheless, the investigated techniques are not
restricted to amateur radio signals and also apply to professional
communications. 

The most common RTTY mode in amateur radio is RTTY 45/170, which uses
a symbol rate of 45.45 baud and a frequency shift of 170 Hz \cite{TheAmericanRadioRelayLeague.2023}.
This mode is frequently used for text message communication on the
shortwave bands, where its simplicity, robustness, and widespread
software support make it a popular mode.

\subsection{Transmitter and Receiver Setup}

\begin{figure}
\begin{centering}
\includegraphics[width=1\columnwidth]{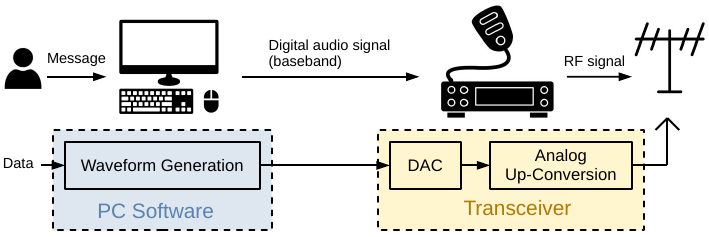}
\par\end{centering}
\caption{\label{fig:Typical-Setup-amateur-radio}Typical setup of an amateur
radio transceiver}
\end{figure}

A typical transmitter setup for digital amateur radio communication
is shown in Figure \ref{fig:Typical-Setup-amateur-radio}. It consists
of a DSP software running on a PC, which generates the digital waveform,
i.e. encodes and modulates the data to be transmitted. The modulated
baseband signal is then passed to the transceiver device as a digital
audio signal. The transceiver typically performs additional DSP, digital-to-analog
conversion and uses analog hardware to up-convert, amplify and filter
the signal before it is transmitted over the antenna.

\subsection{DSP Software for Waveform-Generation}

\begin{figure}
\begin{centering}
\includegraphics[width=1\columnwidth]{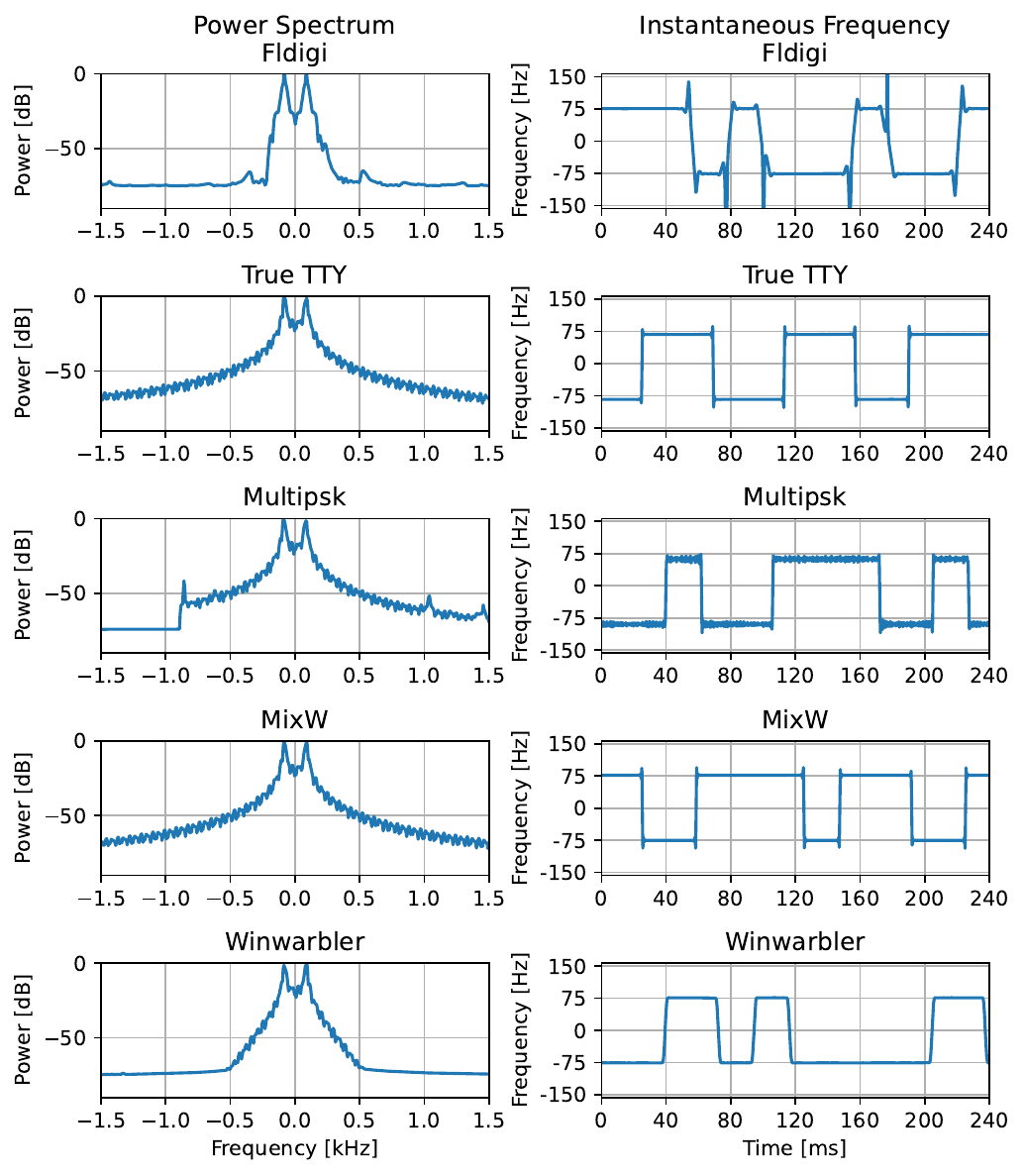}
\par\end{centering}
\caption{\label{fig:Visual-comparison-wide}Visual inspection of the signal's
power spectrum and instantaneous frequency reveals small differences
that could be exploited for fingerprinting. The DSP implementations
differ mainly in spectral skirts, discrete spurious components, and
the shape of frequency transitions. These observations motivate the
use of power spectrum and instantaneous-frequency representations.}
\end{figure}

For the generation of RTTY for amateur radio, different software packages
are available from different manufacturers and authors. The operator
can choose the waveform generation software independently of the transceiver
hardware used. Five different waveform-generation software packages
for RTTY are considered in this paper, each representing one of the
classes in the RF fingerprinting task:

\begin{itemize}
\item Fldigi \cite{fldigi2}: Popular software that supports a multitude
of amateur radio modes (Windows / Linux)
\item TrueTTY \cite{truetty}: focuses on text modes, like RTTY and PSK
(Windows)
\item MultiPSK \cite{multipsk}: wide range of functions, supports a large
number of modes (Windows)
\item MixW \cite{mixw}: supports typically used modes like RTTY, SSTV,
Packet (Windows)
\item Winwarbler \cite{winwarbler}: supports RTTY and PSK modes (Windows)
\end{itemize}
All considered software packages can generate the RTTY 45/170 mode.
However, they use different DSP implementations to generate the signal.
These algorithmic deviations lead to small differences in the generated
signal, which are not well visible in the time domain (IQ). However,
they become more apparent in the power spectrum or the instantaneous
frequency representation. Figure \ref{fig:Visual-comparison-wide}
shows some examples with high signal-to-noise-ratio (SNR) generated
by each of the five software implementations. Some differences in
the spectral shapes can clearly be identified, such as varying shapes
of spectral skirts or spurious emissions. The instantaneous frequency
shows differences in the amount of overshoot, ringing and varying
frequency transitions. It is not straightforward nor required to trace
the distinctive signal properties to specific aspects of the DSP implementations.
Instead, the differences can be automatically extracted from exemplary
data by the training algorithms employed for machine learning.

For the monitoring receiver, it is assumed that the presence of an
RTTY signal is already detected, e.g. using a classifier that can
detect RTTY modes, such as \cite{scholl2025largescaleclassificationshortwavecommunication}.
After that, the proposed method for RF fingerprinting can be applied
to recognize the transmitter by its waveform generation software,
as shown in Figure \ref{fig:hardware-setup}.

\begin{figure}
\begin{centering}
\includegraphics[width=1\columnwidth]{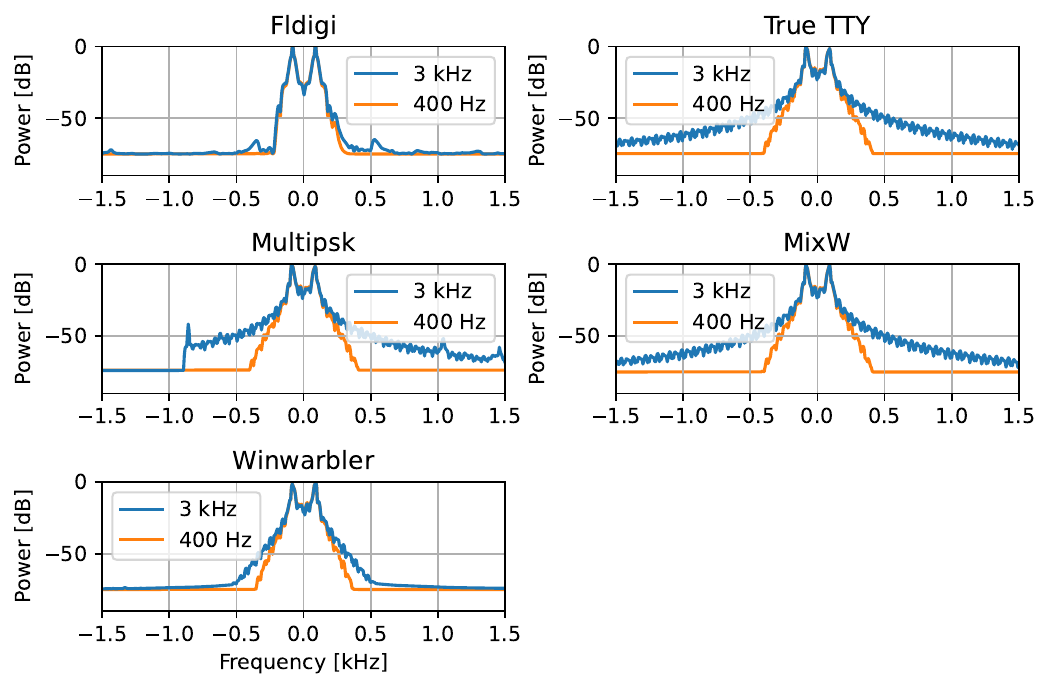}
\par\end{centering}
\caption{\label{fig:spectrum_passband_comparison}Comparison between the spectra
for the 3-kHz and the 400-Hz passband setting. The 400-Hz passband
suppresses some of the implementation-dependent spectral components
outside the nominal RTTY bandwidth, reducing the information available
to the classifier.}
\end{figure}

Two basic cases are considered in the following that use different
RF bandwidths for the transmitted signal and the receiver chain:
\begin{itemize}
\item 400-Hz passband setting: The transmit and monitoring receiver bandwidth
matches that of the RTTY signal.
\item 3-kHz passband setting: The transmit and monitoring receiver bandwidth
is larger than the RTTY signal bandwidth, which also allows the recognition
of out-of-band signal components.
\end{itemize}
The influence of the passband setting on the spectrum is shown in
Figure \ref{fig:spectrum_passband_comparison}.

\subsection{Machine Learning Approach}

\begin{figure}
\begin{centering}
\includegraphics[width=1\columnwidth]{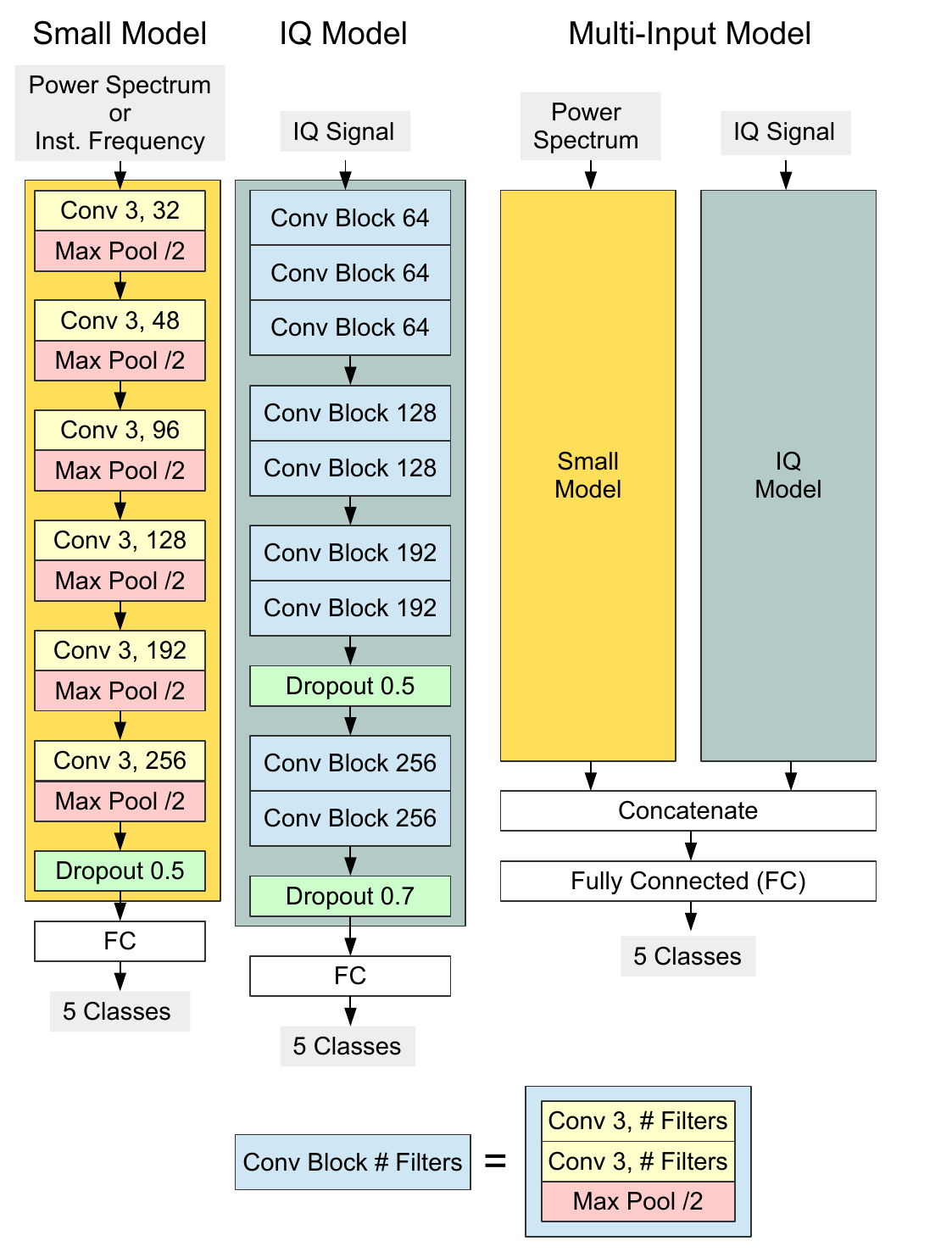}
\par\end{centering}
\caption{\label{fig:Considered-Models}Considered CNN-based models for the
considered input signal representations}
\end{figure}

Algorithms for RF fingerprinting need to extract small differences
from the signal. For that purpose, powerful algorithms are required
that are able to recognize faint patterns in complex signals. Modern
machine learning algorithms, such as neural networks, are well suited
for this task. They learn from a large amount of example data and
are able to classify signals based on subtle properties. Learning
can take place either from raw IQ data or using signal representations
that amplify certain signal characteristics. This paper investigates
four different signal representations for RF fingerprinting:
\begin{itemize}
\item Raw IQ data: complex-valued RF signal without any preprocessing
\item Power spectrum using a 1024-point FFT, highlights spectral properties
\item Instantaneous frequency, highlights properties of frequency transitions
\item A combination of raw IQ data and its power spectrum
\end{itemize}
For each signal representation, a separate neural network model is
used as depicted in Figure \ref{fig:Considered-Models}. The considered
models are CNNs that consist of a combination of convolutional, pooling
and fully connected layers. All convolutional layers use a kernel
size of 3 and ReLU activation functions. Dropout is used for regularization
to prevent overfitting. 

Since the power spectrum and instantaneous frequency highlight the
distinctive signal properties clearly, simple models are sufficient
to exploit the present signal characteristics. This work uses a 13-layer
CNN composed of convolutional and pooling layers, and a final fully
connected softmax layer for class extraction. In contrast to that,
raw IQ signals require more processing to extract distinctive features,
therefore a more complex CNN with 28 layers is used. The combination
of power spectrum and IQ signals is processed with a multi-input model
composed of the IQ and power-spectrum model (without the fully connected
layer) in a parallel arrangement. The power spectrum model extracts
spectral features, while the IQ model extracts the features from the
time domain representation. The output features of both branches are
fused by concatenation and used as input to a dense layer, which extracts
the classes.

\subsection{Training Data and Augmentations}

\begin{figure}
\begin{centering}
\includegraphics[width=1\columnwidth]{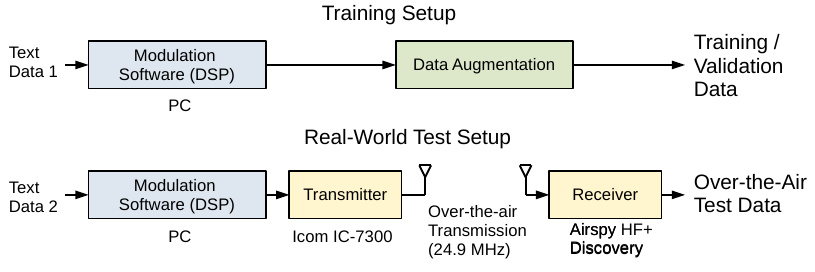}
\par\end{centering}
\caption{\label{fig:Generation-of-training}Generation of training data and
real-world over-the-air test data}

\end{figure}

The training examples are generated using software simulation: First,
signal examples are generated by each of the five waveform-generation
software packages, modulating random text. Second, these clean signals
are then augmented by various RF-related distortions to model real-world
effects as depicted in Figure \ref{fig:Generation-of-training}. A
careful selection of these effects increases the robustness of the
trained model for real-world operation. These signal augmentations
are applied randomly on each data sample:
\begin{itemize}
\item Random frequency offsets from -100 to +100 Hz
\item Random phase offsets
\item Random sample rate deviations up to 1\%
\item Random SNR range from -10 to +50 dB
\item Additive white Gaussian noise (AWGN) with optional impulsive noise
\cite{Scholl.2022}
\item Watterson fading channels according to ITU 1487 \cite{ITUR.052000}
\end{itemize}
Further details on the applied RF signal augmentations can be found
in \cite{Scholl.2022}.

Each training data sample consists of 2048 IQ samples with a sample
rate of 4 kHz and thus has a duration of approximately 0.5 s. For
the power spectrum representation, the FFT window (Hann) covers 1024
samples with an overlap of 50 \% and averages over successive windows.
The power spectrum is represented on a logarithmic scale, i.e. in
dB.

Two datasets have been generated that correspond to the considered
settings of 400-Hz and 3-kHz passband. Each dataset contains 50,000
training examples and 10,000 examples for validation. The datasets
are publicly available from \url{https://panoradio-sdr.de/machine-learning-dataset-for-rf-fingerprinting/}.

\subsection{Model Training}

All models are trained used Adam backpropagation and the categorical
cross-entropy loss. A basic learning-rate scheduling is applied, that
reduces the learning rate when the validation accuracy plateaus. The
models are trained until the validation accuracy converges without
overfitting. This has been achieved in 30 to 50 epochs, depending
on the input representation. For each scenario, i.e. input representation
and passband setting, the average accuracy of three independent training
runs is considered as performance metric.

\subsection{Over-the-Air Tests}

In addition to validation of the neural network on synthetic data,
real-world data is used to evaluate the performance of the system
under more realistic conditions. For that purpose, the waveforms generated
by the software packages are transmitted by a transceiver and recorded
over the air as shown in Figure \ref{fig:Generation-of-training}.
The transmitter is an IC-7300 shortwave transceiver, which transmits
the signal in the shortwave band at 24.9 MHz (12 m amateur radio band).
The receiver is an Airspy HF+ Discovery software defined radio. Note
that the choice of the transmitter and receiver hardware does not
bias the classifier training, because this setup is only used for
generating test data and not training data. 

Over-the-air tests have been conducted using 400-Hz and 3-kHz passband
settings. For each case the transmitter and receiver were set to pass
the specified bandwidth, 3 kHz or 400 Hz. The transmitted text data
is different from the text used for the training data. Since for the
analysis a high SNR is insightful, the transmitter and receiver were
located close to each other to obtain high-SNR recordings. The SNR
is varied by changing the transmit power and the distance between
transmitter and receiver, no artificial augmentations were applied
to the received signals. In total, the over-the-air test data consists
of 19,796 signal examples with a SNR ranging from -10 to 50 dB.

\subsection{Classification Results}

The results show the accuracy of the trained models for the real-world
over-the-air test data. The models for each of the four representations,
raw IQ, power spectrum, instantaneous frequency and the combination
of IQ and power spectrum have been evaluated dependent on the signal
SNR. 

\subsubsection{3-kHz Passband Setting}

\begin{figure}
\begin{centering}
\includegraphics[width=1\columnwidth]{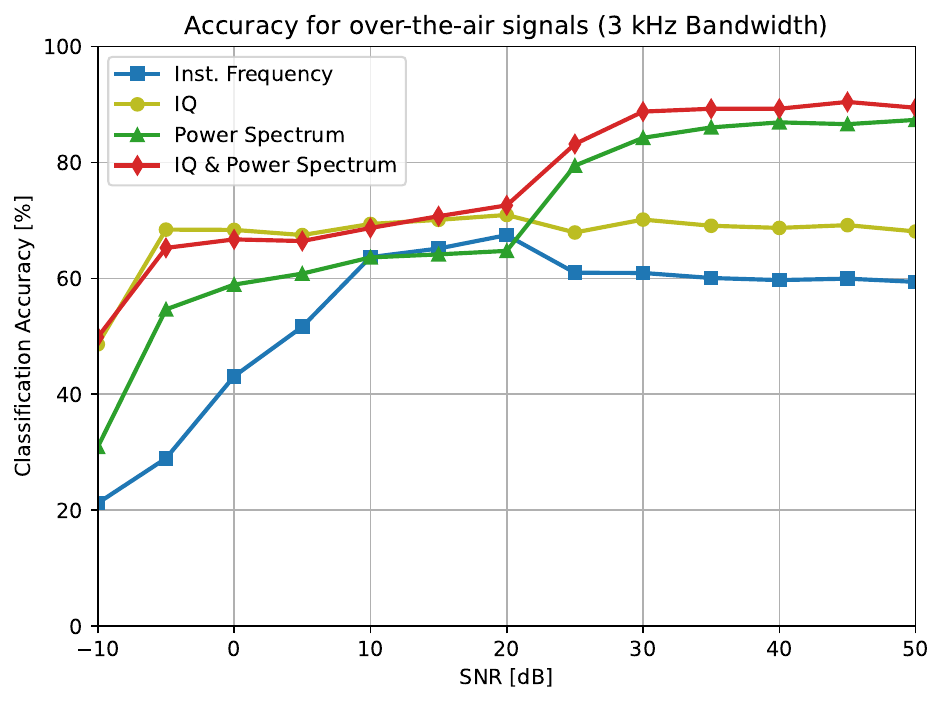}
\par\end{centering}
\caption{\label{fig:results_acc_vs_snr_wideband}Classification accuracy for
the 3-kHz passband setting}

\end{figure}

Figure \ref{fig:results_acc_vs_snr_wideband} shows the accuracy over
SNR for the wider 3-kHz passband setting. The instantaneous frequency
is the least accurate representation and achieves only 60 \% accuracy
even for high SNR values. In contrast, the power spectrum provides
very good results of almost 90 \% at high SNR above 20 dB, while the
raw IQ data performs best for the lower SNR region achieving around
70 \%. The multi-input model using IQ and power spectrum combines
the good low-SNR performance of the IQ data with the high accuracy
of the power spectrum in the high-SNR region and provides the best
overall accuracy.

This shows that the weak spectral features highlighted by the power
spectrum can provide a large advantage, but they are only recognizable
at very high SNRs. These are presumably spectral skirts or weak spurious
emissions that are visible in Figure \ref{fig:Visual-comparison-wide}.
From the IQ data, features could be extracted that are robust to noise
and are still well present for small SNR values below -5 dB.

\begin{figure*}
\begin{centering}
\includegraphics[width=2\columnwidth]{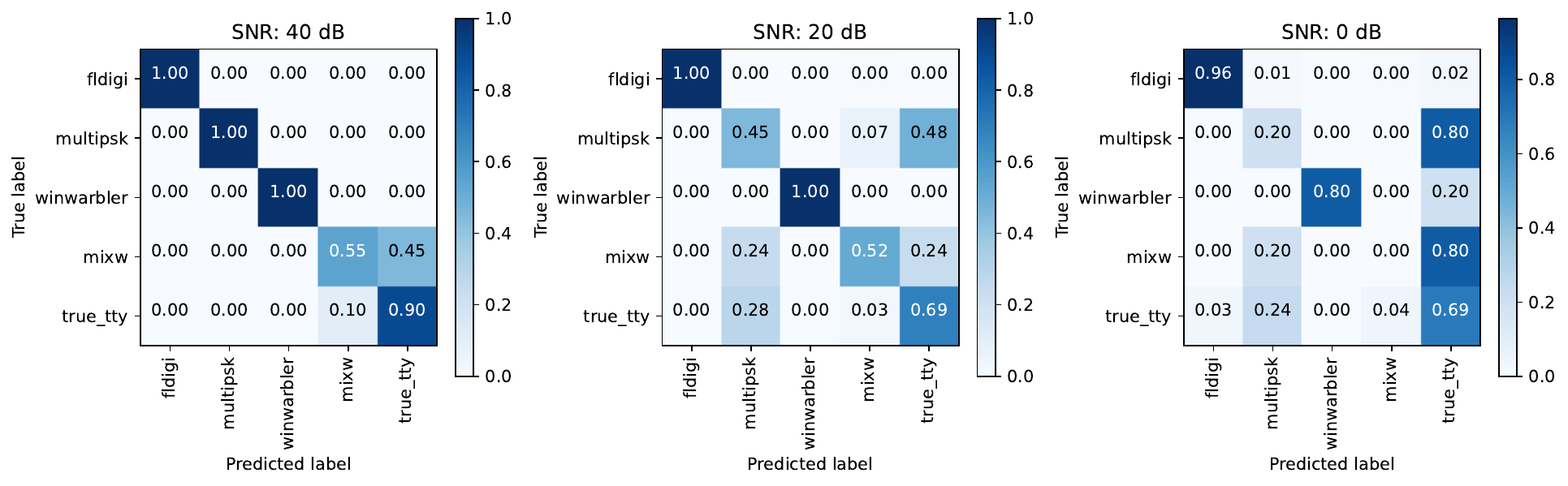}
\par\end{centering}
\caption{\label{fig:results_cm_wide}Confusion matrices for different SNRs
for the 3-kHz passband setting for the multi-input model (IQ and power
spectrum)}
\end{figure*}

Figure \ref{fig:results_cm_wide} shows the confusion matrices for
the multi-input model. For 40 dB SNR, Fldigi, Multipsk and Winwarbler
can be separated very well, while minor confusions occur between MixW
and TrueTTY. For smaller SNRs, Fldigi and Winwarbler remain distinguishable,
while the other three classes are prone to confusions. This highlights
the benefit of a high SNR in RF fingerprinting, which enabled the
highest accuracy of 90~\% for an SNR of +30 dB and above.

\subsubsection{400-Hz Passband Setting}

\begin{figure}
\begin{centering}
\includegraphics[width=1\columnwidth]{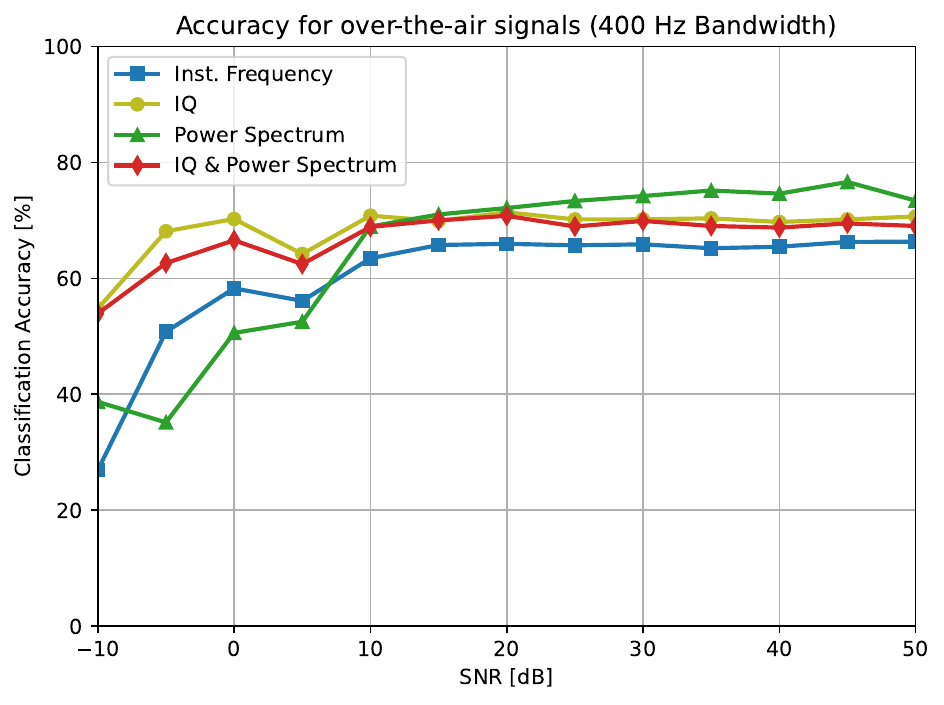}
\par\end{centering}
\caption{\label{fig:results_acc_vs_snr_narrowband}Classification accuracy
for the 400-Hz passband setting}
\end{figure}

\begin{figure*}
\begin{centering}
\includegraphics[width=2\columnwidth]{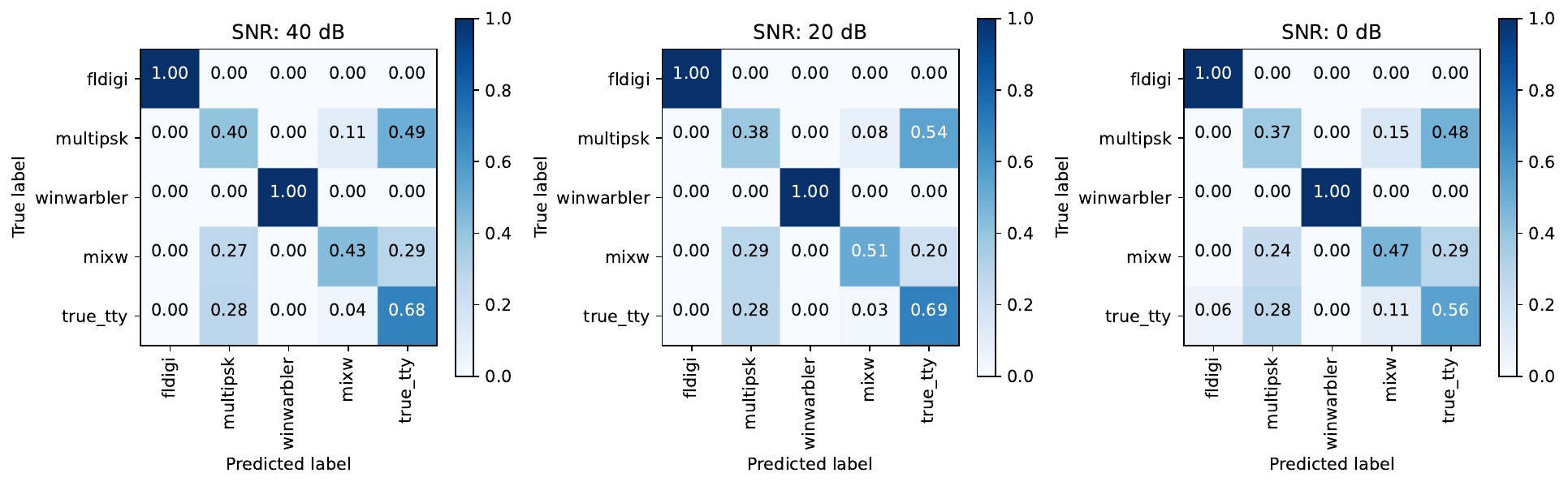}
\par\end{centering}
\caption{\label{fig:results_cm_narrow}Confusion matrices for different SNRs
for the 400-Hz passband setting for the IQ-based model}
\end{figure*}

The results for the 400-Hz passband scenario are shown in Figure \ref{fig:results_acc_vs_snr_narrowband}.
It can be seen that the reduction in bandwidth reduces the accuracy
considerably to around 70 \%, even for high SNR. The 400-Hz passband
benefits less from high SNR. For the 400-Hz passband setting, the
differences between the signal representations are smaller. Here,
the power spectrum provides a slightly better performance than the
other representations at high SNR, while raw IQ data is again superior
at low-SNR as it was the case with the 3-kHz passband setting. The
multi-input model of IQ and power spectrum does not provide an improvement
here. A possible reason is that the narrower passband limits the benefit
of feature fusion. The confusion matrices are shown in Figure \ref{fig:results_cm_narrow}
for the IQ model, which is considered as the best solution here due
to its superior performance at low and medium SNR.

\subsubsection{Results Summary}

The results for this case study show that for SEI it is advantageous
to have high SNR or excess bandwidth. The best results are achieved
when both conditions are fulfilled. In addition, it can be seen that
some classes are highly similar and thus harder to distinguish (e.g.
Multipsk, MixW, TrueTTY), while other classes (Fldigi, Winwarbler)
can be recognized well even under low-SNR and limited passband bandwidth.

\section{\label{sec:Summary}Conclusion}

The paper has investigated RF fingerprinting for a practical setting
of shortwave communication signals. While in the literature RF fingerprinting
often relies on hardware imperfections, this work exploited differences
in the DSP algorithms implemented in a transmitter. In the considered
practical case study, the influence of five different DSP implementations
for the generation of communication waveforms has been investigated.
IQ and power spectrum have been identified as useful signal representations
for CNN classifiers. The results showed that RF fingerprinting works
particularly well for high SNR and if a wide signal bandwidth can
be observed, for which an accuracy of up to 90 \% has been achieved
with over-the-air test signals. Therefore, DSP-related fingerprinting
can provide an effective complementary source of information for SEI
systems.

\bibliographystyle{IEEEtran}
\bibliography{literature}

\end{document}